\documentclass{PoS}

\newcommand{\nn}{\nonumber}
\newcommand{\ovl}[1]{\overline{#1}}

\newcommand{\p}{\partial}

\newcommand{\pslash}{p\kern-1ex /}
\newcommand{\lslash}{l\kern-1ex /}
\newcommand{\kslash}{k\kern-1ex /}
\newcommand{\dslash}{\p\kern-1.2ex /}
\newcommand{\Dslash}{{\cal D}\kern-1.5ex /}
\newcommand{\Aslash}{A\kern-1.2ex /}

\newcommand{\VEV}[3]{\left\langle #1\left| #2 \right| #3\right\rangle}

\newcommand{\les}{\stackrel{<}{{}_{\sim}}}

\title{Light meson spectrum with $N_f=2$ dynamical overlap fermions}

\ShortTitle{Light meson spectrum with $N_f=2$ dynamical overlap fermions}

\author{
  JLQCD collaboration:
  \speaker{J.~Noaki}$^{,a}$\thanks{E-mail: noaki@post.kek.jp},
  S.~Aoki$^b$, 
  H.~Fukaya$^c$,
  S.~Hashimoto$^{a,d}$
  T.~Kaneko$^{a,d}$, 
  H.~Matsufuru$^a$,
  T.~Onogi$^e$,
  E.~Shintani$^a$,
  N.~Yamada$^{a,d}$
  \vspace*{2mm}
  \\
  \llap{$^a$}
  High Energy Accelerator Research Organization (KEK),
  Tsukuba 305-0801, Japan
  \\
  \llap{$^b$}
  Graduate School of Pure and Applied Sciences,
  University of Tsukuba, Tsukuba 305-8571, Japan
  \\
  \llap{$^c$}
  Theoretical Physics Laboratory, RIKEN, Wako 351-0198, Japan
  \\
  \llap{$^d$}
  School of High Energy Accelerator Science,
  the Graduate University for Advanced Studies (Sokendai),
  Tsukuba 305-0801, Japan
  \\
  \llap{$^e$}
  Yukawa Institute for Theoretical Physics, 
  Kyoto University, Kyoto 606-8502, Japan
}

\abstract{ 
We present numerical simulation of QCD with two dynamical quark flavors
described by the overlap fermion action on 
a $16^3\times 32\times (0.12\ {\rm fm})^4$ lattice. 
We calculate pseudo-scalar masses and decay constants and investigate
their chiral properties. We test the consistency of our data with the 
two-loop chiral perturbation theory predictions, which should also be 
valid at finite lattice spacings because of the exact chiral symmetry, 
including the finite size effects.}

\FullConference{The XXV International Symposium on Lattice Field Theory\\
		 July 30 - August 4 2007\\
		 Regensburg, Germany}
\begin{document}

\section{Introduction}

Thanks to the exact chiral symmetry, lattice simulation with the overlap 
fermions~\cite{Neuberger1998} provides a unique opportunity to 
approach low energy hadron physics.
One of the major issues in this energy region is the consistency between 
QCD and chiral perturbation theory (ChPT). 
With the overlap fermion, the continuum ChPT can be applied without 
modification, that is not true with other fermion formulations that
violate either chiral or flavor symmetry.

In this talk, we present two-flavor QCD simulation using the overlap fermion 
and the Iwasaki gauge action. 
On a $16^3\times 32$ lattice, we generate 10,000 
trajectories~\cite{JLQCD2006_1,JLQCD2006_2,JLQCD2006_3, JLQCD2006_4, 
Matsufuru_rev}
at six different sea quark masses.
We explicitly suppress zero-modes of 
hermitian Wilson-Dirac operator $H_W=\gamma_5D_W$ 
in order to avoid discontinuity of the overlap-Dirac operator 
$D_{\rm ov} = m_0(1-\gamma_5\cdot{\rm sgn}(H_W(-m_0)))$ by introducing
extra Wilson fermions which are irrelevant in the continuum~\cite{JLQCD2006_4}.
As a result, the topological charge of our gauge configurations is fixed. 
We have chosen the trivial topological sector $Q_{\rm top} =0$.

In Section 2, we present the spectrum calculation focusing on new
techniques using eigenmodes of the dirac operator. After making
corrections to our data due to the finite size effects (FSE) in Section 3, 
we carry out chiral extrapolation of the corrected results using the NNLO 
ChPT formulae and compare the results with phenomenological values.

\section{Spectrum calculation}

In advance of the spectrum calculation, we compute and store 50 pairs of the
lowest-lying eignmodes of $D_{\rm ov}$ on each gauge configuration.
With these data, we decompose the quark propagator as
\begin{eqnarray}
 S_q(x|y){}_{\alpha\beta} = \sum_{i=-50}^{50}
  \frac{u^i_\alpha(x)\cdot u^i_\beta(y)}
  {\lambda_i +m_q^0}+ S_q^{\rm high}(x|y){}_{\alpha\beta},\label{LMP}
\end{eqnarray}
where the indices $\alpha,\beta$ represent spin$\times$color and $i$
labels eigenmode. $m_q^0$ is the bare quark mass.
While we use the conventional CG algorithm to obtain the second term 
with the source vector with low mode contributions projected out,
we construct the first term from the eigenvalue $\lambda_i$
and eigenvectors $u^i_\alpha (x)$.
Since it is the low mode contribution which is computationally 
dominating the CG iteration, $S_q^{\rm high}$ can be
obtained with much lower computational cost ($\times 8$ less).

\begin{figure}
 \begin{center}
  \includegraphics[width=8.0cm,clip]{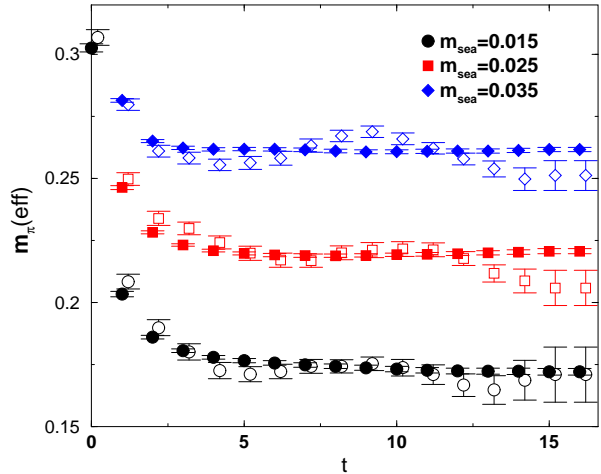}
  \caption{Effective mass from the P-P correlation function with 
  smeared source for the lightest three quark masses 
  $m_{\rm sea} = m_{\rm val}$. Open symbols 
  represent data from conventional correlation function while filled
  ones are from low-mode-averaged correlator.}
  \label{effm}
 \end{center}
\end{figure}

Computation of the meson correlator is done on 500 configurations 
separated by 20 HMC trajectories.
When decomposing quark propagator as in (\ref{LMP}),
a part of the meson correlator is composed purely from the low mode 
contribution.
By replacing this part by its average over the source location
(low-mode-averaging~\cite{DeGrand2004,Giusti2004}), we can improve the data. 
Its advantage is apparent in Figure~\ref{effm} where the effective mass 
of pseudo-scalar correlation functions with (filled symbols) and without 
(open symbols) low-mode-averaging are compared.

We obtain pion mass and decay constant from the pseudo-scalar 
correlators with smeared source and with local source.
By fitting them simultaneously with single exponential functions 
sharing the same mass, 
we obtain the matrix element of the local operator $|\VEV{0}{P}{\pi}|^2$,
from which $f_\pi$ is computed, with a better statistical quality than 
solely using the local-local correlator.
$f_\pi$ is obtained though the axial Ward-Takahashi identity
\begin{eqnarray}
 f_\pi &=&  2m_q^0 \VEV{0}{P}{\pi}/m_\pi^2,
\end{eqnarray}
without further renormalization. Note that we are using the 
$f_\pi = 130$ MeV normalization.

Throughout this calculation, we use $r_0=0.49$ fm as an input. 
Our lattice scale at the chiral limit is 
\begin{eqnarray}
 a &=& 0.1184(12)(11)\ {\rm fm}.
\end{eqnarray}
By using its central value $a^{-1} = 1.67$ GeV, the pion mass data points
are in $290$ MeV $\le m_\pi \le 750$ MeV. 
Our heaviest $m_q^0$ is roughly corresponding to the strange quark mass $m_s$.

We are interested in the chiral extrapolation of $m_\pi^2/m_q$ 
and $f_\pi$, where $m_q$ is the quark mass renormalized by 
the renormalization factor 
\begin{eqnarray}
 Z_m^{\ovl{\rm MS}}(2\ {\rm GeV}) &=& 0.742(12),
\end{eqnarray}
which is obtained non-perturbatively through the RI/MOM scheme on the
lattice~\cite{Martinelli1995}.

\section{Finite size corrections}

\subsection{Standard FSE}

Some of our data points are in the region $m_\pi L< 3.0$ where 
finite size effect could be significant. Therefore, it is important 
to estimate the correction to our data.
To this end, we use the result of analytic calculation by 
Colangelo {\it et al.}~\cite{Colangelo2005}.
Developing from the L\"uscher's formula~\cite{Luscher1986}, 
which relates the mass-shift in a finite box to the amplitude of
scattering of pions, they calculated the finite volume corrections
for $m_\pi$ and $f_\pi$.
Though their calculation includes NNNLO of ChPT for $m_\pi$ and 
NNLO for $f_\pi$, we apply their NNLO results for both quantities.
The NNLO effects depend on the low energy constants (LECs) 
$\bar{l}_{1,2,3,4}$ of the $N_f=2$ ChPT.
At the scale of the physical pion mass $m_\pi^{\rm phys}= 139.6$ MeV,
they are estimated~\cite{Colangelo2001} as
\begin{eqnarray}
 \bar{l}_1^{\rm\ phys} &=& -0.4 \pm 0.6, \label{lphys1}\\
 \bar{l}_2^{\rm\ phys} &=&  4.3 \pm 0.1, \label{lphys2}\\
 \bar{l}_3^{\rm\ phys} &=&  2.9 \pm 2.4, \label{lphys3}\\
 \bar{l}_4^{\rm\ phys} &=&  4.4 \pm 0.2. \label{lphys4}
\end{eqnarray}
We use these phenomenological values to correct our data. The errors
in (\ref{lphys1})--(\ref{lphys4}) are reflected in the results 
assuming a gaussian distribution.
Because the L\"uscher's formula is based on the 
full theory, we can use this result only for the data with 
the same mass for valence and sea quarks. 

\subsection{Correction for fixed $Q_{\rm top}$ }\label{QtopSec}

Since our numerical simulation is done at the topological charge 
fixed to zero, our observables are not free from the artifact.
However, for large enough volume, local quantities such
as hadron mass or matrix elements do not depend on $Q_{\rm top}$.
In fact it is verified by a saddle-point expansion~\cite{Brower2003} that
the difference between the correlation function at fixed topological
charge and that in the $\theta$-vacuum is of ${\cal O}(V^{-1})$. 
By using this relation to pion 
mass and decay constant, it can be shown that the leading correction
is proportional to their second derivative with respect to 
$\theta$ dependence at the saddle point
$\theta_s = iQ_{\rm top}/(V\chi_t)$. $\chi_t$ is the topological 
susceptibility, which is calculated on the same set of configurations 
~\cite{Chiu_writeup}. 
At NLO of ChPT we obtain the corrections
\begin{eqnarray}
 \frac{m_\pi^{Q_{\rm top}=0}}{m_\pi(\theta=0)}
  &=& 1-\frac{1}{16V\chi_t}
  \left[1+
   \left(\frac{m^{\rm tree}_\pi(\theta=0)}{4\pi f}\right)^2
   \left(
    \ln\left(\frac{m^{\rm tree}_\pi(\theta=0)}{m_\pi^{\rm phys}}\right)^2 
    -\bar{l}_3^{\ \rm phys}+1
   \right)
  \right],\nn \\
 \\
 \frac{f_\pi^{Q_{\rm top}=0}}{f_\pi(\theta=0)}
  &=& 1+ \frac{1}{4V\chi_t}
  \left(\frac{m^{\rm tree}_\pi(\theta=0)}{4\pi f}\right)^2
  \left(
   \ln\left(\frac{m^{\rm tree}_\pi(\theta=0)}{m_\pi^{\rm phys}}\right)^2 
   -\bar{l}_4^{\ \rm phys}+1
  \right),
\end{eqnarray}
where $m^{\rm tree}_\pi(\theta)^2 = 2B_0m_q \cos (\theta/N_f)$
is the tree-level $\theta$-dependent pion mass.
The fixed $Q_{\rm top}$ correction starts 
at the tree-level for $m_\pi$ while it does at NLO for $f_\pi$.

\begin{figure}
 \begin{center}
  \includegraphics[width=7.2cm,clip]{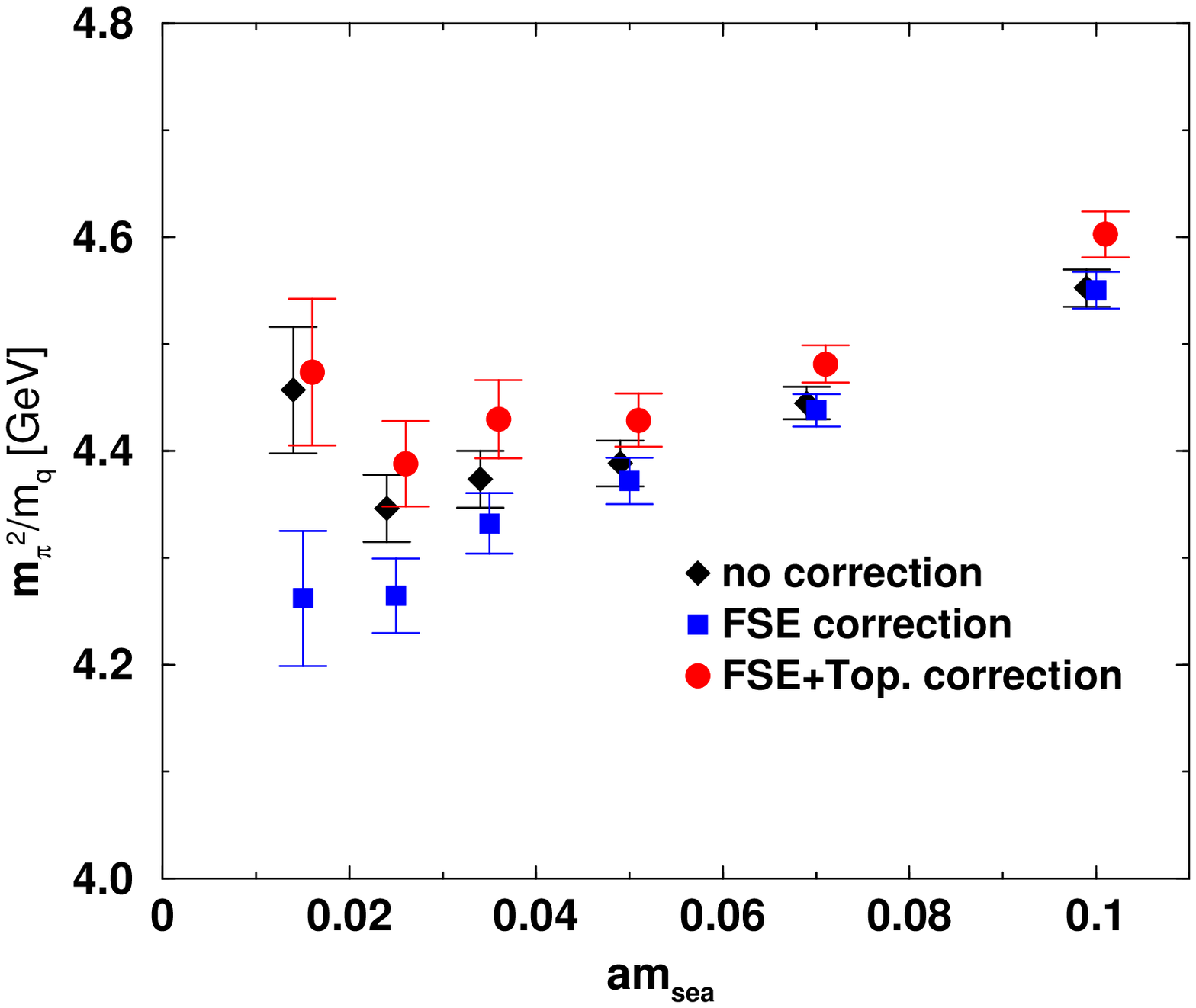}
  \includegraphics[width=7.4cm,clip]{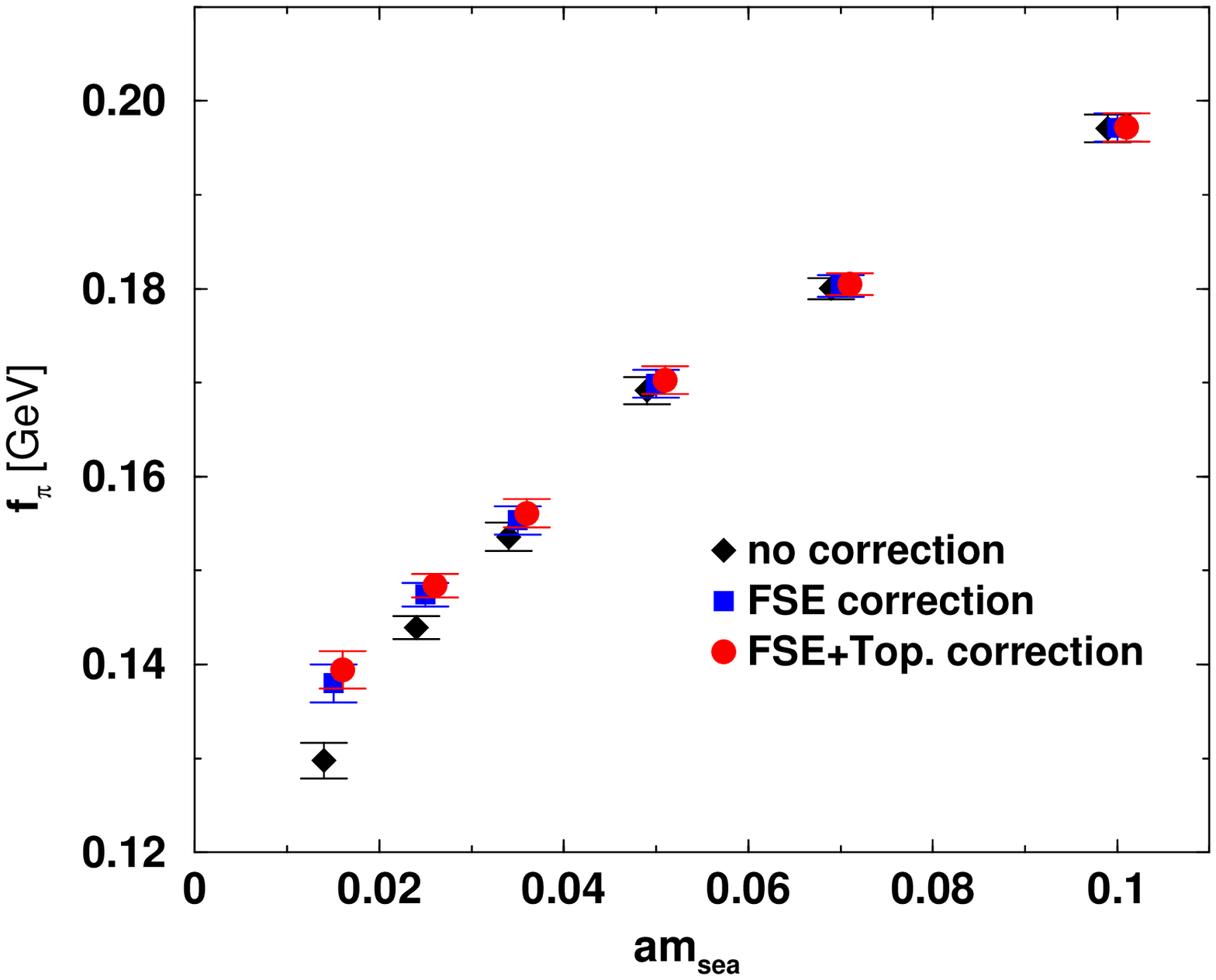} 
  \caption{Results of finite size corrections for $m_\pi^2/m_q$ 
  (left) and $f_\pi$ (right). Data points as a function of $m_{\rm sea}$ 
  are slightly shifted for visibility.}
  \label{FSEres}
 \end{center}
\end{figure}

In Figure~\ref{FSEres}, we illustrate the effects of finite size effects 
for $m_\pi^2/m_q$ (left) and $f_\pi$ (right).
In each panel, the original data (black diamonds) are corrected by the 
standard FSE (blue squares) and further by the fixed topology effect 
(red circles). For the data with the lightest quark mass, the
standard FSE is 4.5 \% and 6.0 \% for $m_\pi^2/m_q$ and
$f_\pi$, respectively.
For $m_\pi^2/m_q$ in small mass region, two finite size 
effects appear with an opposite sign to almost cancel each other.

\section{Chiral extrapolation}

While conventional $N_f=2$ full ChPT expansion formulae are written 
in power of 
$\left(m_\pi^2\right)^{\rm tree}= 2 B_0 m_q$, 
we chose $\xi=(m_\pi/(4\pi f_\pi))^2$ as an expansion parameter. 
This choice enables us to fit the ChPT formulae (almost) independently to 
$m_\pi^2/m_q$ and $f_\pi$ as a function of $\xi$ which is 
a measurable quantity at each quark mass. The NLO formulae are
\begin{eqnarray}
 \left(m_\pi^2/m_q \right)_{\rm NLO}&=& 2B_0 
  \left( 1 +\xi\ln\xi -\bar{l}_3\xi\right),\label{Mchiral_NLO}\\
 \left(f_\pi\right)_{\rm NLO} &=& f
  \left( 1 -2\xi\ln\xi +2\bar{l}_4\xi\right) \label{Fchiral_NLO}
\end{eqnarray}
where $\bar{l}_{3,4}$, $B_0$ and $f$ are to be fitted.
$\bar{l}_{3,4}$ are LECs at the scale $4\pi f$.
Since these two formulae do not share any fit parameter, independent fits
are possible.

We also examine the NNLO formulae which are written as
\begin{eqnarray}
 \left(m_\pi^2/m_q \right)_{\rm NNLO}&=& 2B_0 
  \left[
   1 +\xi\ln\xi +\frac{7}{2}(\xi\ln\xi)^2
   -\bar{l}_3(\xi -9\xi^2\ln\xi) 
   +\left(
     \bar{l}_4 -\frac{4}{3}\left(\tilde{l}^{\rm\ phys}+16\right)
   \right)\xi^2 \ln\xi
  \right]\nn\\
 & &  
  +\alpha_1\xi^2+{\cal O}(\xi^3),\label{Mchiral}\\
 \nn\\
 \left(f_\pi\right)_{\rm NNLO} &=& f
  \left[
   1 -2\xi\ln\xi +5(\xi\ln\xi)^2 
   +2\bar{l}_4(\xi-10\xi^2\ln\xi) 
   +\frac{3}{2}\left(\tilde{l}^{\rm\ phys}+\frac{53}{2}\right)\xi^2\ln\xi
  \right]\nn\\
 & & +\alpha_2\xi^2+{\cal O}(\xi^3),\label{Fchiral}
\end{eqnarray}
where coefficients of the $\xi^2$ term $\alpha_1$ and $\alpha_2$
contain unknown low energy constants associated with the NNLO counterterms.
In carrying out the chiral fit with these formulae, an input
\begin{eqnarray}
 \tilde{l}^{\rm\ phys}\equiv 7\,\bar{l}_1^{\rm\ phys}+8\,\bar{l}_2^{\rm\ phys}
  -15\ln \left(\frac{4\pi f_\pi^{\rm phys}}{m_\pi^{\rm phys}}\right)^2
\end{eqnarray}
is introduced from phenomenological estimates (\ref{lphys1}) 
and (\ref{lphys2}) with $f_\pi^{\rm phys}=130.0$ MeV. 

In our target range $0< \xi \les 0.1$, 
an approximation $\xi^2\ln \xi\approx -2.5 \xi^2$ is numerically 
rather precise.
Therefore, it is reasonable to consider yet another fit ansatz with 
modified NNLO formulae, which we call NNLO'
\begin{eqnarray}
 \left(m_\pi^2/m_q \right)_{\rm NNLO'}&=& 2B_0
  \left[
   1 +\xi\ln\xi +\frac{7}{2}(\xi\ln\xi)^2 -\bar{l}_3\xi\right] 
  +\alpha'_1\xi^2+{\cal O}(\xi^3),\label{Mchiral_NNLO'}\\
 \nn\\
 \left(f_\pi\right)_{\rm NNLO'} &=& f
  \left[
   1 -2\xi\ln\xi +5(\xi\ln\xi)^2 +2\bar{l}_4\xi\right] 
  +\alpha'_2\xi^2+{\cal O}(\xi^3).\label{Fchiral_NNLO'}
\end{eqnarray}

\begin{figure}
 \begin{center}
  \includegraphics[width=7.3cm,clip]{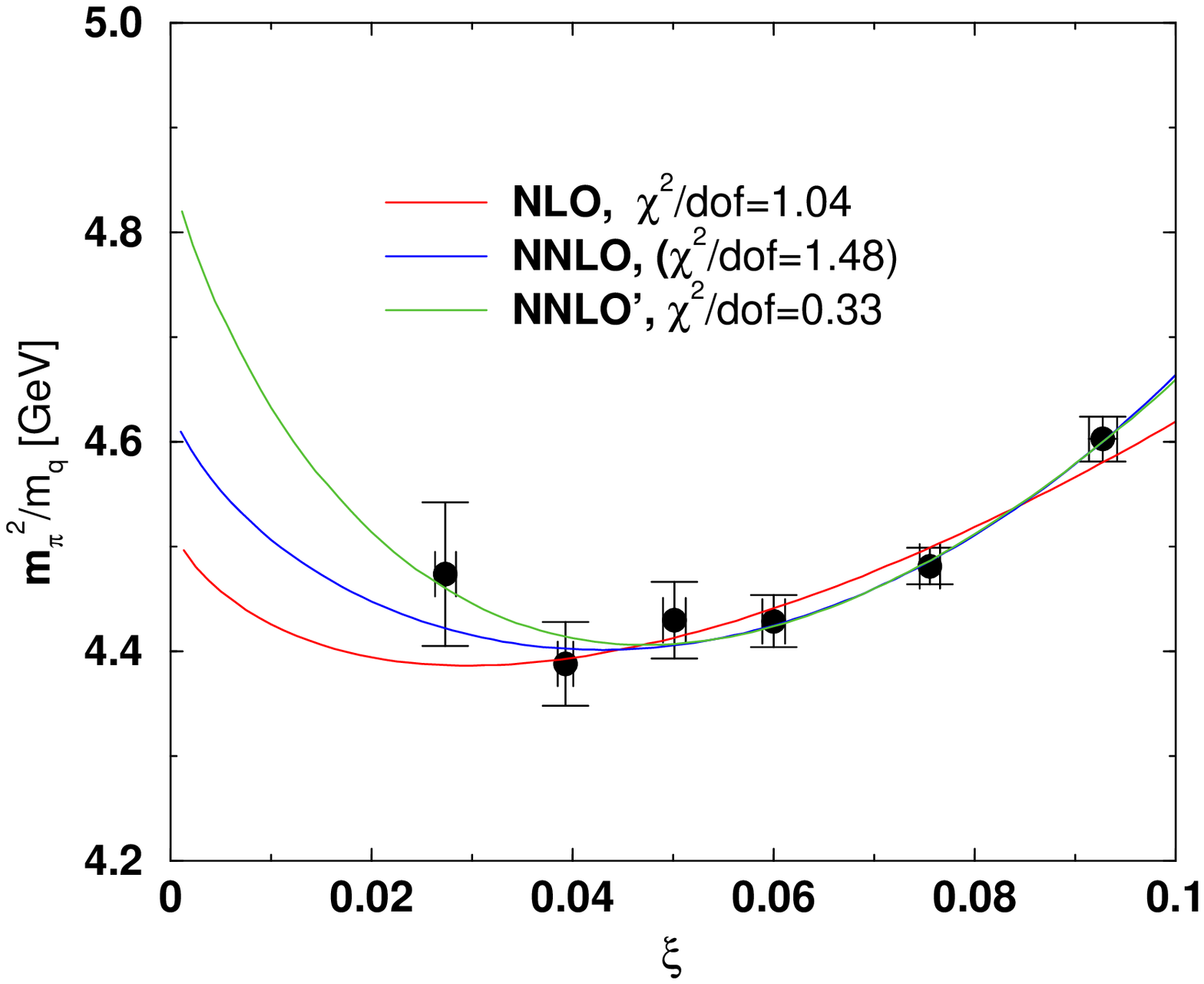} 
  \includegraphics[width=7.5cm,clip]{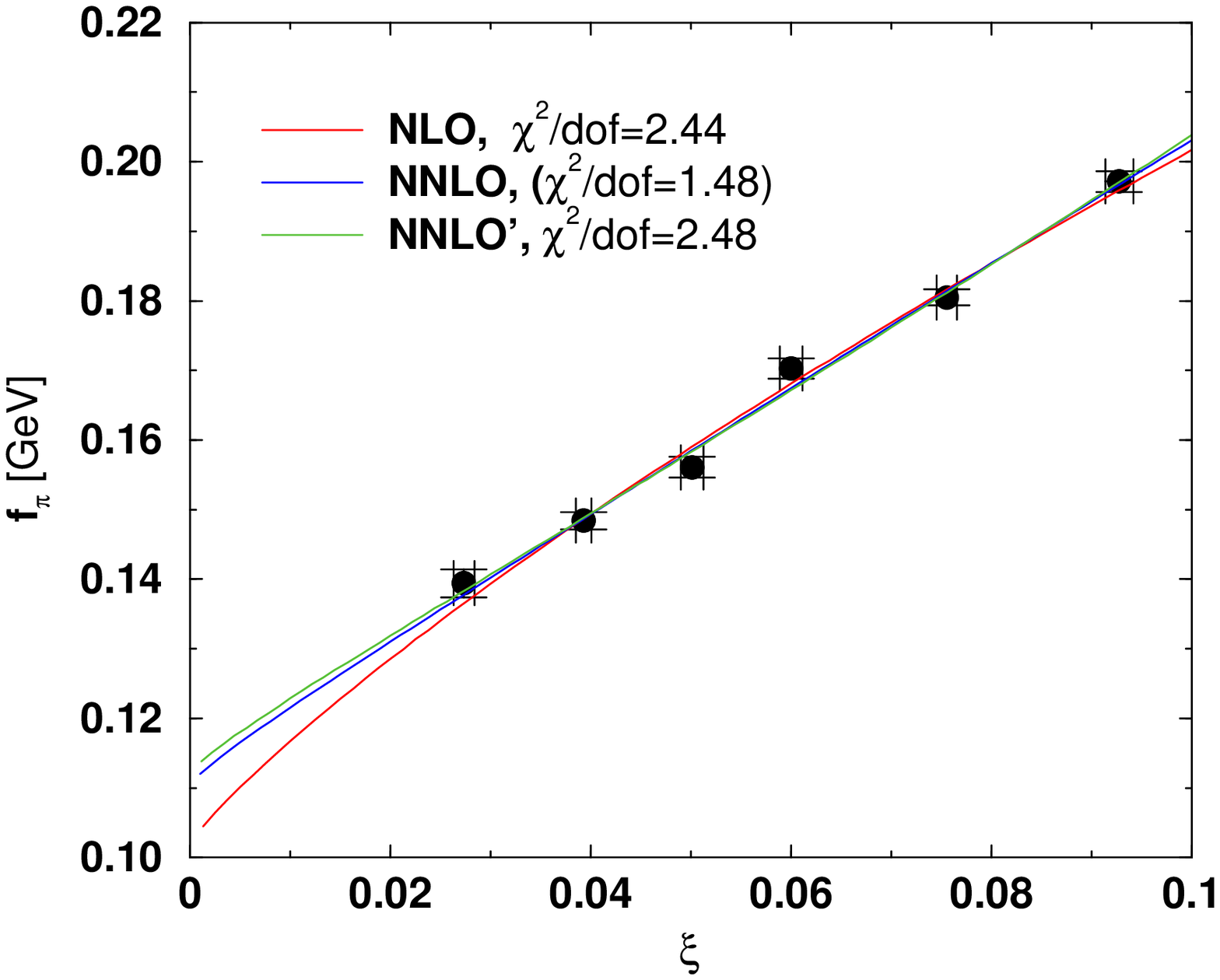} 
  \caption{Results of chiral extrapolation for $m_{\pi}^2/m_q$ [GeV]
 (left) and $f_{\pi}$ [GeV] (right) as a function of $\xi$. 
  For each panel, three kinds of fit (NLO, NNLO and NNLO') by using 
  all data points are shown.}
  \label{XFit}
 \end{center}
\end{figure}

In Figure~\ref{XFit} we show the results of chiral 
fits to the FSE corrected data of $m_\pi^2/m_q$ and $f_\pi$, 
respectively. 
In each panel, we compare NLO (red curves), NNLO (blue) and NNLO' (green).
Note that we cannot compare the quality of the fit with NNLO and others 
by the values of $\chi^2/{\rm dof}$ because we carry out the
simultaneous fit for the former and independent fits for the latter.

\begin{figure}
 \begin{center}
  \hspace{0.5mm}
  \includegraphics[width=6.5cm,clip]{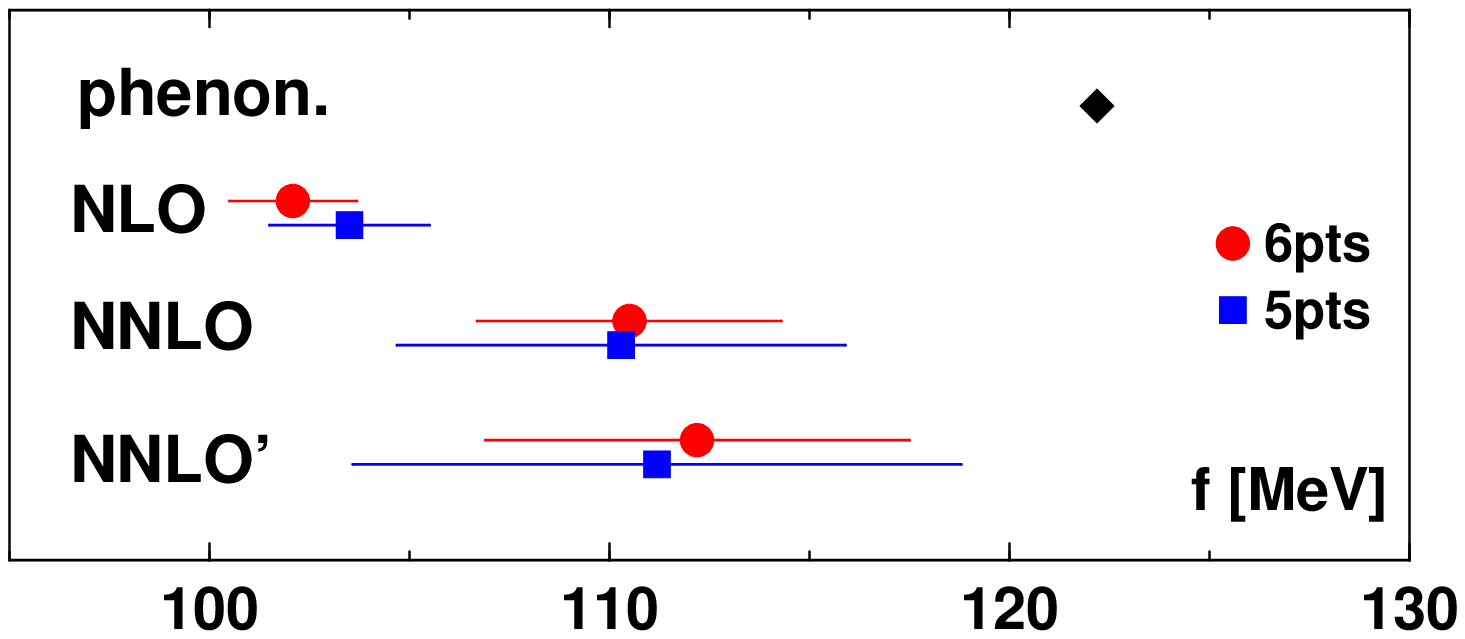} 
  \hspace{0.2mm}
  \includegraphics[width=6.5cm,clip]{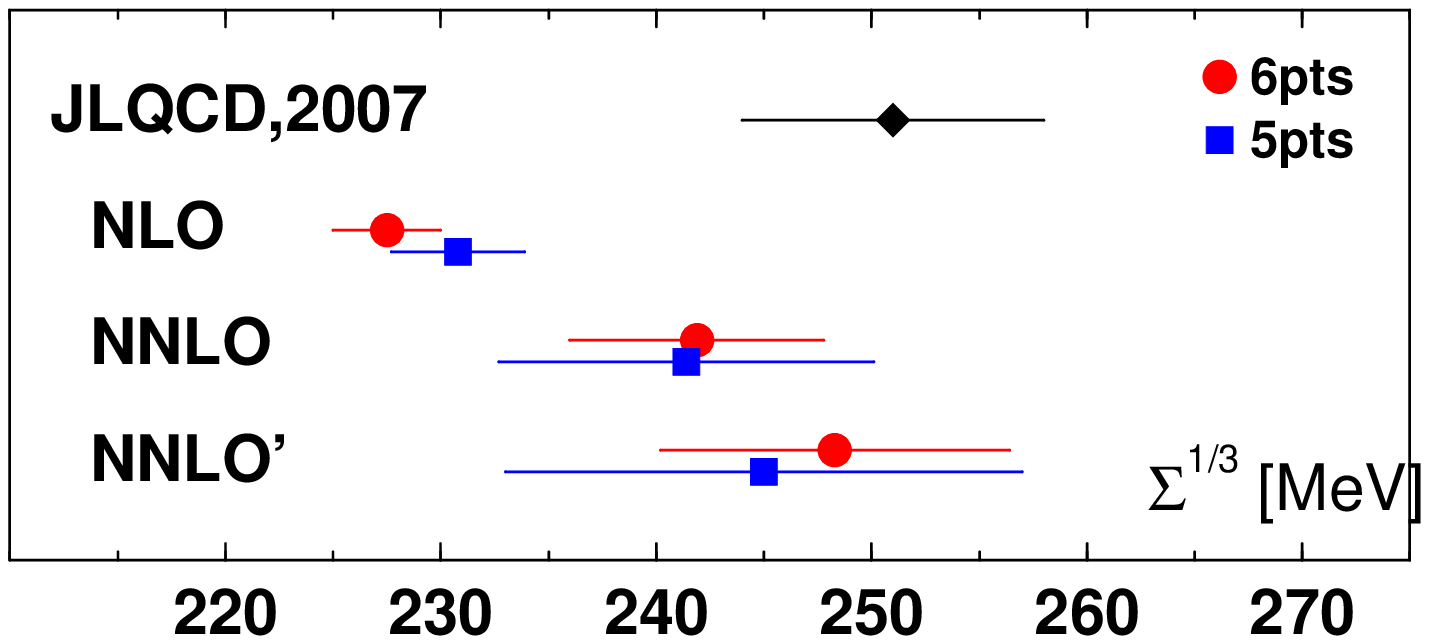} \\
 \end{center}
 \begin{center}
  \includegraphics[width=6.4cm,clip]{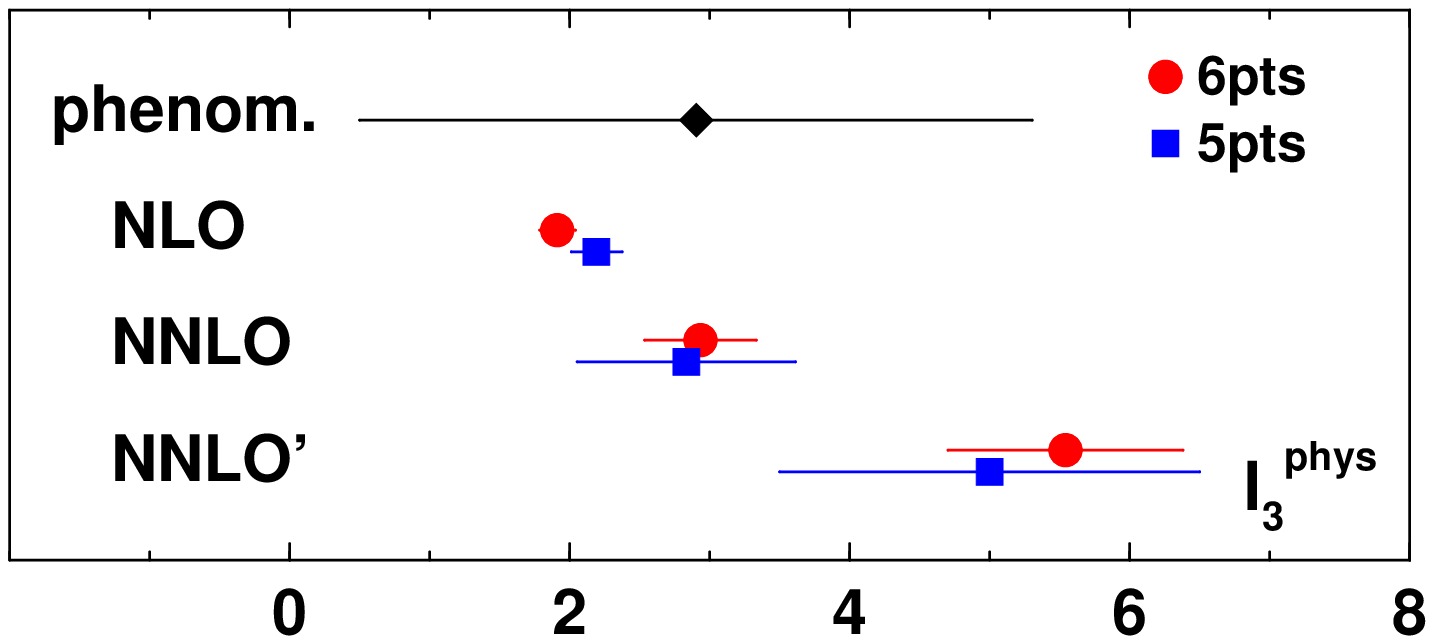} 
  \hspace{1mm}
  \includegraphics[width=6.4cm,clip]{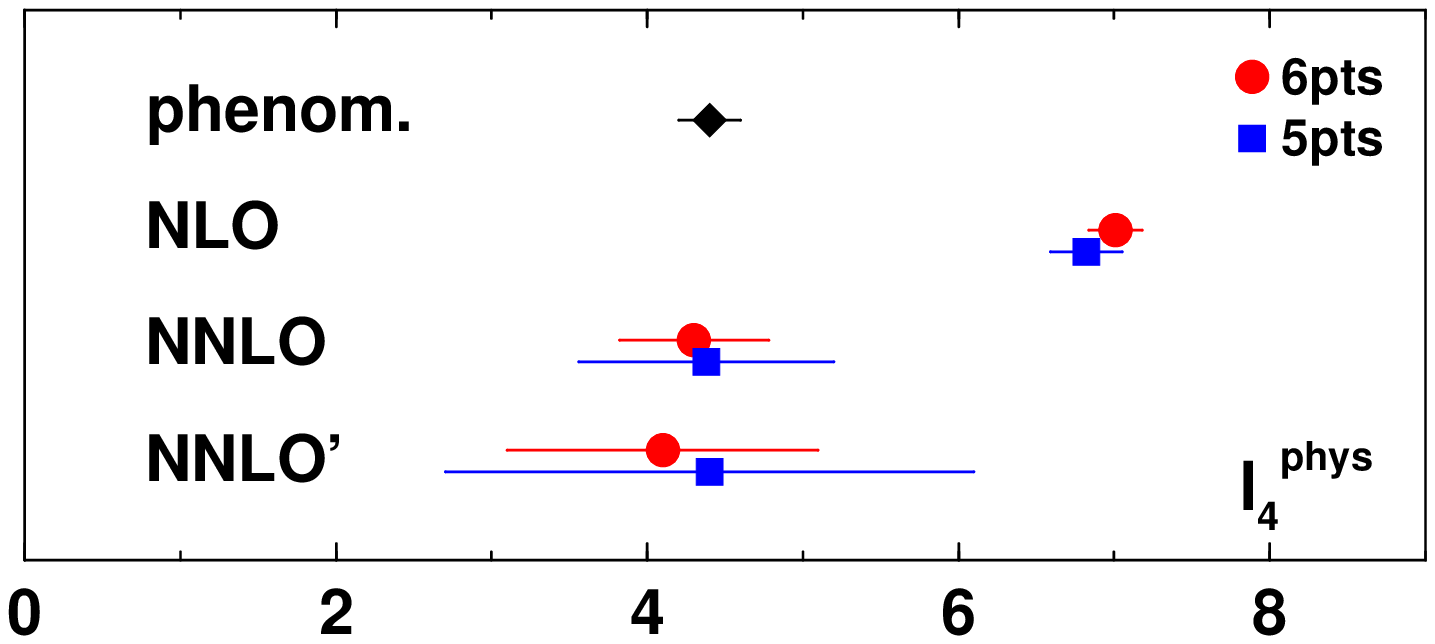} 
  \end{center}these reference points.
 \caption{Comparison of physical quantities obtained from 
 chiral fit ansatz: $f$ (upper left), $\Sigma^{1/3}$ (upper right), 
 $\bar{l}_3^{\rm phys}$ (lower left) and  
 $\bar{l}_4^{\rm phys}$ (lower right). In each panel, red circles and 
 blue squares are  corresponding to fits with 6 and 5 lightest data points, 
 respectively.}
 \label{Results}
\end{figure}

Figure~\ref{Results} compares quantities from our fit: $f$,\ $\Sigma =
B_0\cdot f^2/2$,\ $\bar{l}_3^{\rm phys}$ 
and $\bar{l}_4^{\rm phys}$.
In the figure we also plot phenomenological values of $f$~\cite{Gasser1984}
and $\bar{l}_{3,4}^{\rm \ phys}$ ((\ref{lphys3}) and (\ref{lphys4})) 
as well as the result of our previous calculation
in $\epsilon$-regime~\cite{Fukaya2007}: $\Sigma^{1/3} = 0.251(7)(11)$ GeV.

We see that, for $\bar{l}_3^{\rm phys}$, large error of the 
phenomenological estimate covers all of our results.
For other quantities, results from NLO fit are 
inconsistent with the NNLO results. 
This implies the failure of the NLO formulae to describe the data
in the quark mass region up to $m_s$.
Results from NNLO and NNLO' are consistent with each other and with 
phenomenological estimates.

For the future we are planning to extend the present
analysis to $m_K$ and $f_K$ on the $N_f=2+1$ dynamical lattices 
(see~\cite{HashimotoProc} for the status). Numerical check of the estimate 
of FSE in a bigger volume is also planned.

\vspace*{5mm}
Numerical simulations are performed on Hitachi SR11000 and
IBM System Blue Gene Solution at High Energy Accelerator Research
Organization (KEK) under a support of its Large Scale
Simulation Program (No.~07-16).
This work is supported in part by the Grant-in-Aid of the
Ministry of Education 
(Nos. 
13135204, 
15540251, 
17740171,
18034011, 
18340075, 
18740167,
18840045,
19540286,
19740121,
19740160).

\newcommand{\J}[4]{{#1} {\bf #2} (#3) #4}
\newcommand{\RMP}{Rev.~Mod.~Phys.}
\newcommand{\MPL}{Mod.~Phys.~Lett.}
\newcommand{\IJMP}{Int.~J.~Mod.~Phys.}
\newcommand{\NP}{Nucl.~Phys.}
\newcommand{\NPSup}{Nucl.~Phys.~{\bf B} (Proc.~Suppl.)}
\newcommand{\PL}{Phys.~Lett.}
\newcommand{\PRD}{Phys.~Rev.~D}
\newcommand{\PRL}{Phys.~Rev.~Lett.}
\newcommand{\AP}{Ann.~Phys.}
\newcommand{\CMP}{Commun.~Math.~Phys.}
\newcommand{\CPC}{Comp.~Phys.~Comm.}
\newcommand{\PTP}{Prog. Theor. Phys.}
\newcommand{\Suppl}{Prog. Theor. Phys. Suppl.}
\newcommand{\JHEP}{JHEP}

\end{document}